\documentclass[11pt]{amsart}
\usepackage{graphicx, amssymb}

\newtheorem{thm}{Theorem}[section]

\newtheorem{prop}[thm]{Proposition}
\theoremstyle{definition}
\newtheorem{defn}[thm]{Definition}

\theoremstyle{remark}

\newtheorem{exa}[thm]{Example}
\numberwithin{equation}{section}

\newcommand{\Real}{\mathbb R}

\newcommand{\such}{\ | \ }
\newcommand{\prob}{\mathbb{P}}

\newcommand{\qprob}{\mathbb{Q}}
\newcommand{\expec}{\mathbb{E}}

\newcommand{\cov}{\mathrm{cov}}

\newcommand{\basis}{(\Omega, \, \mathcal{F}, \, \mathbf{F}, \, \prob)}
\newcommand{\filtration}{\mathbf{F} = \pare{\mathcal{F}_t}_{t \in [0, T]}}
\newcommand{\F}{\mathcal{F}}

\newcommand{\ud}{\mathrm d}
\newcommand{\inner}[2]{\left \langle #1 , #2 \right \rangle}

\newcommand{\pare}[1]{\left(#1\right)}
\newcommand{\bra}[1]{\left[#1\right]}
\newcommand{\dbra}[1]{[\kern-0.15em[ #1 ]\kern-0.15em]}
\newcommand{\dbraco}[1]{[\kern-0.15em[ #1 [\kern-0.15em[}
\newcommand{\dbraoc}[1]{]\kern-0.15em] #1 ]\kern-0.15em]}

\newcommand{\fone}{\mathbf{1}}
\newcommand{\bF}{\mathbf{F}}

\newcommand{\pro}{\mathsf{P}}
\newcommand{\qpro}{\mathsf{Q}}

\begin{document}

\title[Stochastic Discount Factors]{Stochastic Discount Factors}%

\author{Constantinos Kardaras}%
\address{Constantinos Kardaras, Mathematics and Statistics Department, Boston University, 111 Cummington Street, Boston, MA 02215, USA.}%
\email{kardaras@bu.edu}%

\keywords{Arbitrage, asset pricing, (local) martingales, risk
premium, state-price densities, utility indifference pricing}%

\date{\today}%
\begin{abstract}
The valuation process that economic agents undergo for investments
with uncertain payoff typically depends on their statistical views
on possible future outcomes, their attitudes toward risk, and,
of course, the payoff structure itself. Yields vary across different
investment opportunities and their interrelations are difficult to
explain. For the same agent, a different discounting factor has to be
used for every separate valuation occasion. If, however, one is ready
to accept discounting that varies randomly with the possible outcomes,
and therefore accepts the concept of a stochastic discount factor,
then an economically consistent theory can be developed. Asset
valuation becomes a matter of randomly discounting payoffs under
different states of nature and weighing them according to the agent's
probability structure. The advantages of this approach are obvious,
since a single discounting mechanism suffices to describe how any
asset is priced by the agent.
\end{abstract}

\maketitle

\setcounter{section}{-1}

\section{Introduction}

Within active and liquid financial markets, economic agents are able to make investment decisions. Capital is allocated today in exchange for some future income stream. If there is no uncertainty regarding the future payoff of an investment opportunity, the yield that will be asked on the investment will equal the risk-free interest rate prevailing for the time period covering the time of investment until the time of the payoff. However, in the presence of any payoff uncertainty when undertaking an investment venture, economic agents will typically ask for \emph{risk compensation}, and thus for some \emph{investment-specific} yield, which will discount the expected future payoff stream. The yields, that particular agents ask for, depend both on their statistical views on possible future outcomes, as well as their attitudes towards risk.

Yields vary across different investment opportunities and their interrelations are difficult to explain. For the \emph{same} agent, a different discounting factor has to be used for every separate valuation occasion. If one, however, is ready to accept discounting that varies \emph{randomly} with the possible outcomes, and therefore accepts the concept of a \textsl{stochastic} discount factor, then a very economically consistent theory can be developed. Asset valuation becomes a matter of randomly discounting payoffs under different states of nature and weighing them according to the agent's probability structure. The advantages of this approach are obvious, since a \emph{single} discounting mechanism suffices to describe how \emph{any} asset is priced by the agent.

\smallskip

In Section \ref{sec: SDF_discrete}, the theory of stochastic discount factors for the discrete-time, discrete-state case is presented. In this framework, all the concepts and ideas are introduced naked of any technical complications.  Later, in Section \ref{sec: SDF_Ito}, the theory of stochastic discount factors is presented in the more practical case of It\^o-process models.

\section{Stochastic Discount Factors in Discrete Probability Spaces} \label{sec: SDF_discrete}

We start by introducing all relevant ideas in a very simple one-time-period framework and finite states-of-the-world. There are plenty of textbooks with great exposition on these, as well as related, themes, such as \cite{citeulike:975777} or the first chapters of \cite{MR1422250} --- see also \cite{MR2169807} for the general state-space case.

\subsection{The set-up}

Consider a very simplistic example of an economy, where there are only two dates of interest, represented by times $t = 0$ (today) and $t = T$ (financial planning horizon). There are several states-of-nature possible at time $T$ and for the time being these are represented as a \emph{finite} set $\Omega$. Only one $\omega \in \Omega$ will be revealed at time $T$, but this is not known in advance today.

In the market, there is a \emph{baseline} asset with a price process $S^0 = (S^0)_{t = 0, T}$. Here, $S^0_0$ is a strictly positive constant and $S^0_T (\omega) > 0$ for all $\omega \in \Omega$. The process $\beta := S^0_0 / S^0$ is called the \textsl{deflator}. It is customary to regard this baseline asset as \emph{riskless}, providing a simple annualized interest rate $r \in \Real_+$ for investment from today to time $T$; in this case, $S^0_0 = 1$ and $S^0_T = 1 + r T$. This viewpoint is \emph{not} adapted here, since it is unnecessary.

Together with the baseline asset, there exist $d$ other liquid \emph{traded} assets whose prices $S^i_0$, $i = 1, \ldots, d$ today are known constants, but the prices $S^i_T$, $i = 1, \ldots, d$, at day $T$ depend on the outcome $\omega \in \Omega$, i.e., they are random variables.

\subsection{Agent portfolio selection via expected utility maximization}

Consider an economic agent in the market as described above. Faced with inherent uncertainty, the agent postulates some likelihood on the possible outcomes, modeled via a \textsl{probability mass} $\pro : \Omega \mapsto [0,1]$ with $\sum_{\omega \in \Omega} \pro[\omega] = 1$. This gives rise to a probability $\prob$ on the subsets of $\Omega$ defined via $\prob[A] = \sum_{\omega \in A} \pro(\omega)$ for all $A \subseteq \Omega$. This probability can either be \emph{subjective}, i.e., coming from views that are agent-specific, or \emph{historical}, i.e., arising from statistical considerations via some estimation procedure.

Economic agents act in the market and optimally invest in order to maximize their satisfaction. Each agent has some \emph{preference structure} on the possible future random payoffs that is represented here via the \textsl{expected utility} paradigm.\footnote{One can impose natural conditions on preference relations defined on the set of all possible outcomes that will lead to numerical representation of the preference relationship via expected utility maximization. This was axiomatized in \cite{MR2316805} --- see also Chapter 2 of \cite{MR2169807} for a nice exposition.} There exists a \emph{continuously differentiable}, \emph{increasing} and \emph{strictly concave} function $U : \Real \mapsto \Real$, such that the agent will prefer a random payoff $\xi: \Omega \mapsto \Real$ from another random payoff $\zeta: \Omega \mapsto \Real$ at time $T$ if and only if $\expec^\prob[U(\zeta)] \leq \expec^\prob[U(\xi)]$, where $\expec^\prob$ denotes expectation with respect to the probability $\prob$.

Starting with capital $x \in \Real$, an economic agent chooses at day zero a strategy $\theta \equiv (\theta^1, \ldots, \theta^d) \in \Real^d$, where $\theta^j$ denotes the units from the $j^{th}$ asset held in the portfolio. What remains, $x - \sum_{i=1}^d \theta^i S^i_0$, is invested in the baseline asset. If $X^{(x, \theta)}$ is the wealth generated starting from capital $x$ and investing according to $\theta$, then $X^{(x, \theta)}_0 = x$ and
\begin{equation} \label{eq: wealth disc}
X^{(x; \, \theta)}_T = \left( x - \sum_{i=1}^d \theta^i S^i_0 \right) \frac{S_T^0}{S_0^0} + \sum_{i=1}^d \theta^i S_T^i = x \frac{S_T^0}{S_0^0}   + \sum_{i=1}^d \theta^i \left( S_T^i - \frac{S_T^0}{S_0^0} S_0^i  \right),
\end{equation}
or in deflated terms $\beta_T X^{(x; \, \theta)}_T = x + \sum_{i=1}^d \theta^i \left( \beta_T S_T^i - S_0^i \right)$.
The agent's objective is to choose a strategy in such a way as to \emph{maximize expected utility}, i.e.,
find $\theta_*$ such that
\begin{equation} \label{eq: UMP}
\expec^\prob \bra{U \pare{X^{(x; \, \theta_*)}_T}} = \sup_{\theta \in \Real^d}
\expec^\prob \bra{U \pare{X^{(x; \, \theta)}_T}}.
\end{equation}
The above problem will indeed have a solution if and only if no arbitrages exist in the market. By definition, an \emph{arbitrage} is a wealth generated by some $\theta \in \Real^d$ such that $\prob [X^{(x; \, \theta)}_T \geq 0] = 1$ and $\prob [X^{(x; \, \theta)}_T > 0] > 0$. It is easy to see that arbitrages exist in the market if and only if $\sup_{\theta \in \Real^d} \expec^\prob [U (X^{(x; \, \theta)}_T)]$ is \emph{not} attained by some $\theta_* \in \Real^d$. Assuming then the \textsl{No-Arbitrage} (NA) condition, concavity of the function $\Real^d \ni \theta \mapsto \expec^\prob [U (X^{(x; \, \theta)}_T)]$ will imply that the first-order conditions
\[
\frac{\partial}{ \partial \theta^i} \Big{|}_{\theta = \theta_*}
\expec^\prob \bra{U \pare{X^{(x; \, \theta)}_T}} = 0, \quad \text{for all } i = 1, \ldots, d,
\]
will provide the solution $\theta_*$ to the problem.
Since the expectation is just a finite sum, the differential operator can pass inside, and then the first-order conditions for optimality are
\begin{equation} \label{eq: 1st-order for util opt}
0 = \expec^\prob \left[ \frac{\partial}{\partial \theta^i}
\Big{|}_{\theta = \theta_*} U \pare{X^{(x; \, \theta)}_T} \right] =
\expec^\prob \bra{ U' \pare{ X^{(x; \, \theta_*)}_T } \left( S_T^i - \frac{S_T^0}{S_0^0} S_0^i  \right) }, \quad i=1, \ldots, d.
\end{equation}
The above is a non-linear system of $d$ equations to be solved for $d$ unknowns $(\theta_*^1, \ldots, \theta_*^d)$. Under (NA), the system \eqref{eq: 1st-order for util opt} has a solution $\theta_*$. Actually, under a trivial non-degeneracy condition in the market the solution is unique; even if the optimal strategy $\theta_*$ is not unique, strict concavity of $U$ implies that the optimal wealth $X^{(x; \, \theta_*)}_T$ generated is unique.

A little bit of algebra on (\ref{eq: 1st-order for util opt}) gives, for all $i =1, \ldots, d$,
\begin{equation} \label{eq: disc factor asset valua discrete}
S^i_0 = \expec^\prob \left[ Y_T
S^i_T \right], \text{ where } Y_T \, := \,  \frac{ U' \pare{X^{(x; \, \theta_*)}_T}}{\expec^\prob \bra{(S_T^0 / S_0^0) U' \pare{ X^{(x; \, \theta_*)}_T}}}
\end{equation}
Observe that since $U$ is continuously differentiable and strictly increasing, $U'$ is a strictly positive function, and therefore $\prob[Y_T > 0] = 1$.  Also, \eqref{eq: disc factor asset valua discrete} trivially also holds for $i = 0$. Note that the random variable $Y_T$ that was obtained above depends on the utility function $U$, the probability $\prob$, as well as on the initial capital $x \in \Real$.

\begin{defn}
In the model described above, a process $Y = (Y_t)_{t = 0, T}$ will be called a \textsl{stochastic discount factor} if $\prob[Y_0 =1, Y_T > 0] = 1$ and $S^i_0 = \expec^\prob \left[ Y_T
S^i_T \right]$ for all $i=0, \dots, d$.
\end{defn}

If $Y$ is a stochastic discount factor,  using \eqref{eq: wealth disc}  one can actually show that
\begin{equation} \label{eq: disc factor wealth mart discrete}
\expec^\prob \left[ Y_T
X^{(x; \, \theta)}_T \right] = x, \quad \text{for all } x \in \Real \text{ and } \theta \in \Real^d.
\end{equation}
In other words, the process $Y X^{(x; \, \theta)}$ is a $\prob$-martingale for all $x \in \Real$ and $\theta \in \Real^d$.

\subsection{Connection with risk-neutral valuation}

Since $\expec^\prob[S^0_T Y_T] = S_0^0 > 0$, we can define a probability mass $\qpro$ by requiring that $\qpro (\omega) = \left( S^0_T (\omega) /  S^0_0  \right) Y_T(\omega) \pro(\omega)$, which defines a probability $\qprob$ on subsets of $\Omega$ in the obvious way. Observe that, for any $A \subseteq \Omega$, $\qprob [A] > 0$ if and only if $\prob [A] > 0$; we say that the probabilities $\prob$ and $\qprob$ are \emph{equivalent} and we denote by $\qprob \sim \prob$.
Now, rewrite \eqref{eq: disc factor asset valua discrete} as
\begin{equation} \label{eq: EMM in util opt}
S^i_0  = \expec^{\qprob} \left[  \beta_T S^i_T \right], \quad \text{for all } i = 0, \ldots, d.
\end{equation}

A probability $\qprob$, equivalent to $\prob$, with the property prescribed in \eqref{eq: EMM in util opt} is called \textsl{risk-neutral} or an \textsl{equivalent martingale measure}.  In this simple framework, stochastic discount factors and risk-neutral probabilities are in one-to-one correspondence. In fact, more can be said.

\begin{thm} [Fundamental Theorem of Asset Pricing] \label{thm: FTAP}
In the discrete model as described previously, the following three conditions are equivalent.
\begin{enumerate}
  \item There are no arbitrage opportunities.
  \item A stochastic discount factor exists.
  \item A risk neutral probability measure exists.
\end{enumerate}
\end{thm}

The Fundamental Theorem of Asset Pricing was first formulated in \cite{MR0429063} and it took twenty years to reach a very general version of it in general semimartingale models that are beyond the scope of our treatment here. The interested reader can check the monograph \cite{MR2200584}, where the history of the Theorem and all proofs is presented.

%

\subsection{The important case of the logarithm} \label{subsec: log-utility}

The most well-studied case of utility on the real line is $U(x) = \log (x)$, both because of its computational simplicity and for the theoretical value that it has. Since the logarithmic function is only defined on the strictly positive real line, it does not completely fall in the aforementioned framework, but it is easy to see that the described theory is still valid.

Consider an economic agent with logarithmic utility that starts with initial capital $x = 1$. Call $X^* = X^{(1; \, \theta_*)}$ the optimal wealth corresponding to log-utility maximization. The fact that $U'(x) = 1 / x$ allows to define a stochastic discount factor $Y^*$ via $Y^*_0 = 1$ and
\[
Y^*_T \, = \, \frac{1} {X^*_T \expec^\prob \bra{1 / (\beta_T X^*_T)}}
\]
From $\expec^\prob [ Y^*_T X^*_T ] = 1$ it follows that $\expec^\prob [1 / (\beta_T X^*_T)] = 1$ and therefore $Y^* = 1 / X^*$. This simple relationship between the log-optimal wealth and the stochastic discount factor that is induced by it will be one of the keys to characterize existence of stochastic discount factors in more complicated models and their relationship with absence of free lunches. It will find good use in the next Section \ref{sec: SDF_Ito} for the case of models using It\^o processes.

\subsection{Arbitrage-free prices}

For a claim with random payoff $H_T$ at time $T$, an \textsl{arbitrage-free} price $H_0$ is a price at time zero such that the extended market that consists of the original traded assets with asset prices $S^i$, $i=0, \ldots, d$, augmented by the new claim, remains arbitrage-free. If the claim is \textsl{perfectly replicable}, i.e., if there exists $x \in \Real$ and $\theta \in \Real^d$ such that $X^{(x; \, \theta)}_T = H_T$, it is easily seen that the \emph{unique} arbitrage-free price for the claim is $x$. However, it is frequently the case that a newly-introduced claim in not perfectly replicable using the existing liquid assets. In that case, there exist more than one arbitrage-free price for the claim; actually, the set of all the possible arbitrage-free prices is $\{ \expec^\prob [Y_T H_T] \such Y \text{ is a stochastic discount factor}\}$. To see this, first pick a stochastic discount factor $Y_T$ and set $H_0 = \expec[Y_T H_T]$; then, $Y$ remains a stochastic discount factor for the extended market, which therefore does not allow for any arbitrage opportunities. Conversely, if $H_0$ is an arbitrage-free price for the new claim, we know from Theorem \ref{thm: FTAP} that there exists a stochastic discount factor $Y$ for the extended market, which satisfies $H_0 = \expec[Y_T H_T]$ and is trivially a stochastic discount factor for the original market. The result we just mentioned gives justice to the appellation ``Fundamental Theorem of Asset Pricing'' for Theorem \ref{thm: FTAP}.

\subsection{Utility Indifference Pricing}

Suppose that a new claim promising some random payoff at time $T$ is issued. Depending on the claim's present traded price, an economic agent might be inclined to take a long or short position --- this will depending on whether the agent considers the market price low or high, respectively. There does exist a market-price-level of the claim that will make the agent indifferent between longing or shorting an \emph{infinitesimal}\footnote{We stress ``infinitesimal'' because when the portfolio holdings of the agent change, the indifference prices also change; thus, for large sales or buys that will considerably change the portfolio structure, there might appear incentive, that was not there before, to sell or buy the asset.} amount of
asset. This price level is called \textsl{indifference price}. In the context of claim valuation, utility indifference prices have been introduced in \cite{MR1491376}\footnote{For this reason, utility indifference prices are sometimes referred to as ``Davis prices''.}, but have been widely used previously in the Economics science. Indifference prices depend on the particular agent's views, preferences, as well as portfolio structure, and should not be confused with market prices, which are established using the forces of supply and demand.

Since the discussed preference structures are based on
\emph{expected utility}, it makes sense to try and understand quantitatively how \textsl{utility indifference prices} are formed. Under the present set-up, consider a claim with random payoff $H_T$ at time $T$. The question we wish to answer is: what is the indifference price
$H_0$ of this claim today for an economic agent.

For the time being, let $H_0$ be any price set by the market for the claim. The agent will invest in
the risky assets and will hold $\theta$ units of them, as well as the new claim, taking a position of $\epsilon$ units. Then, the agent's terminal payoff is
\[
X_T^{(x; \, \theta, \epsilon)} \, := \, X^{(x; \, \theta)}_T + \epsilon \pare{ H_T - \frac{S_T^0}{S_0^0} H_0 }.
\]
The agent will again maximize expected utility, i.e., will invest $(\theta_*, \epsilon_*) \in \Real^d \times \Real$ such that
\begin{equation} \label{eq: UMP+V}
\expec^\prob \bra{ U\pare{ X_T^{(x; \, \theta_*, \epsilon_*)}}} = \sup_{(\theta, \epsilon) \in \Real^d \times \Real} \expec^\prob \bra{ U\pare{ X_T^{(x; \, \theta, \epsilon)}}}.
\end{equation}
If $H_0$ is the agent's indifference price, it must be that $\epsilon_* = 0$ in the above maximization problem; then, the agent's optimal decision regarding the claim would be not to buy or sell any units of it. In particular, the concave function $\Real \ni  \epsilon \mapsto \expec^\prob \bra{ U\pare{ X_T^{(x; \, \theta, \epsilon)}}}$
should achieve its maximum at $\epsilon = 0$. First-order conditions give that $H_0$ is the agent's indifference price if
\[
0 = \frac{\partial}{\partial \epsilon} \Big|_{\epsilon = 0}  \expec^\prob \bra{ U\pare{ X_T^{(x; \, \theta_*, \epsilon)}} } = \expec^\prob \bra{ U' \pare{ X_T^{(x; \, \theta_*, 0)}}  \left( X_T - \frac{S_T^0}{S_0^0} X_0 \right)}
\]
A remark is in order before writing down the indifference-pricing formula. The strategy $\theta_*$ that has been appearing above represents the optimal holding in the liquid traded assets when \emph{all} assets \emph{and} the claim are available --- it is \emph{not} in general the agent's optimal asset holdings if the claim was \emph{not} around. Nevertheless, \emph{if} the solution of problem (\ref{eq: UMP+V}) is such that the optimal holdings in the claim are $\epsilon_* = 0$, \emph{then} $\theta_*$ are also the agent's optimal asset holdings if there was no claim to begin with. In other words, if $\epsilon_* = 0$, $X_T^{(x; \, \theta_*, 0)}$ is exactly the same quantity $X_T^{(x; \, \theta_*)}$ that appears in \eqref{eq: 1st-order for util opt}. Remembering the definition of the stochastic discount factor $Y_T$ of \eqref{eq: disc factor asset valua discrete}, we can write
\[
H_0 = \expec^\prob \left[ Y_T
H_T \right]
\]
It is important to observe that $Y_T$ depends on a lot of things: namely, the probability $\prob$, the utility $U$ and the initial capital $x$, but \emph{not} on the particular claim
to be valued. Thus, we need only one evaluation of the stochastic discount factor and we can use it to find indifference prices with respect to all sorts of
different claims.

\subsection{State price densities} \label{subsec: state price densities}
For a fixed $\omega \in \Omega$, consider an \textsl{Arrow-Debreau} security that pays off a unit of account at time $T$ if the state-of-nature is $\omega$, and nothing otherwise. The indifference price of this security for the economic agent is $p(\omega) := Y(\omega) \pro(\omega)$. Since $Y$ appears as the density of the ``state price'' $p$ with respect to the probability $\prob$, stochastic discount factors are also termed \textsl{state price densities} in the literature. For two states-of-the-nature $\omega$ and $\omega'$ of $\Omega$ such that $Y(\omega) < Y(\omega')$, an agent that uses the stochastic discount factor $Y$ considers $\omega'$ a more unfavorable state than $\omega$ and is inclined to pay more for insurance against adverse market movements.

\subsection{Comparison with real-world valuation}

Only for the purposes of what is presented here, assume that $S^0_0 = 1$ and $S^0_T = 1 + r T$ for some $r \in \Real_+$. Let $Y$ be a stochastic discount factor; then, we have $1 = S^0_0 = \expec^\prob [Y_T S^0_T] = (1 + r T) \expec^\prob [Y_T]$. Pick then any claim with random payoff $H_T$ at time $T$ and use $H_0 = \expec^\prob [Y_T H_T]$ to write
\begin{equation} \label{eq: risk-neutral vs stoch disc fact}
H_0 =  \frac{1}{1 + r T} \expec^\prob [H_T] + \cov^\prob (Y_T, \, H_T),
\end{equation}
where $\cov^\prob (\cdot, \cdot)$ is used to denote \textsl{covariance} of two random variables with respect to $\prob$. The first term $(1 + r T)^{-1} \expec^\prob [H_T]$ of the above formula describes ``real world'' valuation for an agent who would be neutral under his views $\prob$ in facing the risk coming from the random payoff $H_T$. This risk-neutral attitude is absent usually: agents require compensation for the risk they undertake, or might even feel inclined to pay more for a security that will insure them in cases of unfavorable outcomes. This is exactly mirrored by the correction factor $\cov^\prob (Y_T, \, H_T)$ appearing in \eqref{eq: risk-neutral vs stoch disc fact}. If the covariance of $Y_T$ and $H_T$ is negative, the claim tends to pay more when $Y_T$ is low. By the discussion in \S \ref{subsec: state price densities}, this means that the payoff will be high in states that are not greatly feared by the agent, who will therefore be inclined to pay \emph{less} than what the real-world valuation gives. On the contrary, if the covariance of $Y_T$ and $H_T$ is positive, $H_T$ will pay off higher in dangerous states-of-nature for the agent (where $Y_T$ is also high), and the agent's indifference price will be \emph{higher} than real-world valuation.

\section{Stochastic Discount Factors for It\^o Processes} \label{sec: SDF_Ito}

\subsection{The model}

Uncertainty is modeled via a probability space $\basis$, where $\filtration$ is a filtration representing the flow of information. The market consists of a \emph{locally} riskless savings account whose price process $S^0$ satisfies $S^0_0 > 0$ and
\[
\frac{\ud S^0_t}{S^0_t} = r_t \ud t, \quad t \in [0, T],
\]
for some $\bF$-adapted, positive \textsl{short rate} process $r = (r_t)_{t \in \Real}$. It is obvious that $S^0_t = S^0_0 \exp (\int_0^t r_u \ud u)$ for $t \in [0, T]$. We define the \textsl{deflator} $\beta$ via
\[
\beta_t = \frac{S^0_0}{S^0_t} = \exp \pare{ - \int_0^t r_u \ud u}, \quad t \in [0, T].
\]
The movement of $d$ risky assets will be modeled via It\^o processes:
\[
\frac{\ud S^i_t}{S^i_t} \, = \, b^i_t \ud t + \inner{\sigma^{\cdot i}_t}{\ud W_t}, \quad t \in \Real_+, \quad i=1, \ldots, d.
\]
Here, $b = (b^1, \ldots, b^d)$ is the $\bF$-adapted $d$-dimensional process of \textsl{appreciation rates}, $W = (W^1, \ldots, W^m)$ is an $m$-dimensional $\prob$-\textsl{Brownian motion} representing the \emph{sources of uncertainty} in the market, and $\inner{\cdot}{\cdot}$ denotes the usual inner product notation: $\inner{\sigma^{\cdot i}_t}{\ud W_t} = \sum_{j=1}^m \sigma^{j i}_t \ud W^j_t$ where $(\sigma^{ji})_{1 \leq j \leq m, \, 1 \leq i \leq d}$ if the $\bF$-adapted $(m \times d)$-matrix-valued process whose entry $\sigma^{ji}_t$ represents the impact of the $j$th source of uncertainty on the $i$th asset at time $t \in [0, T]$. With ``$\top$'' denoting transposition, $c := \sigma^\top \sigma$ is the $d \times d$ local \textsl{covariation matrix}. To avoid degeneracies in the market, it is required that $c_t$ has full rank for all $t \in [0, T]$, $\prob$-almost surely. This implies in particular that $d \leq m$ --- there are more sources of uncertainty in the market than are liquid assets to hedge away the uncertainty risk. Model of this sort are classical in the Quantitative Finance literature --- see for example \cite{MR1640352}.

\begin{defn}
A \textsl{risk premium} is any $m$-dimensional, $\bF$-adapted process $\lambda$ satisfying $\sigma^\top \lambda = b - r \fone$, where $\fone$ is the $d$-dimensional vector with all unit entries.
\end{defn}

The terminology ``risk premium'' is better explained for the case $ d = m = 1$; then $\lambda = (b - r) / \sigma$ is the premium that is asked by investors for the risk associated with the (only) source of uncertainty. In the general case, $\lambda^j$ can be interpreted as the premium required for the risk associated with the $j$th source of uncertainty, represented by the Brownian motion $W^j$. In \emph{incomplete} markets, when $d < m$, Proposition \ref{prop: market prices of risk} shows all the different choices for $\lambda$. Each choice will parametrize the different risk attitudes of different investors. In other words, risk premia characterize the possible stochastic discount factors, as shall be revealed in Theorem \ref{thm: structure of SDFs}.

If $m = d$, the equation $\sigma^\top \lambda = b - r \fone$ has only one solution: $\lambda^* = \sigma c^{-1} (b - r \fone) $. If $d < m$ there are many solutions, but they can be characterized using easy linear algebra.
\begin{prop} \label{prop: market prices of risk}
The risk premia are \emph{exactly}
all processes of the form $\lambda = \lambda^* + \kappa$, where $\lambda^* := \sigma c^{-1} (b - r \fone)$ and $\kappa$ is \emph{any} adapted process with $\sigma^\top \kappa = 0$.
\end{prop}
If $\lambda = \lambda^* + \kappa$ in the notation of Proposition \ref{prop: market prices of risk}, then $\inner{\lambda^*}{\kappa} = (b - r \fone)^\top c^{-1} \sigma^\top \kappa =  0$. Then, $| \lambda |^2 = |\lambda^*|^2 + | \kappa |^2$, where $|\lambda^*|^2 = \inner{b - r \fone}{c^{-1} (b - r \fone)}$.

\subsection{Stochastic discount factors} The usual way of obtaining stochastic discount factors in continuous time is through risk-neutral measures. The Fundamental Theorem of Asset Pricing in the present It\^o process setting states that absence of Free Lunches with Vanishing Risk\footnote{Free Lunches with Vanishing Risk is the suitable generalization of the notion of Arbitrages in order to get a version of the Fundamental Theorem of Asset Pricing in continuous time. The reader is referred to \cite{MR2200584}.} is equivalent to the existence of a probability $\qprob \sim \prob$ such that $\beta S^i$ is (only) a \emph{local} $\qprob$-martingale for all $i = 0, \ldots, d$. (For the definition of local martingales, check for example \cite{MR1121940}.) In that case, by defining $Y$ via $Y_t = \beta_t (\ud \qprob / \ud \prob) |_{\F_t}$, $Y S^i$ is a local $\prob$-martingale for all $i = 0, \ldots, d$. The last property will be taken here as the \emph{definition} of a stochastic discount factor.

\begin{defn} \label{defn: stoch disc fact Ito}
Consider the above It\^o process set-up. A stochastic process $Y$ will be called a \textsl{stochastic discount factor} if
\begin{itemize}
  \item $Y_0 = 1$ and $Y_T > 0$, $\prob$-almost surely.
  \item $Y S^i$ is a local $\prob$-martingale for all $i =0, 1, \ldots, d$.
\end{itemize}
\end{defn}

In the case where $Y S^0$ is an actual martingale, i.e., $\expec^\prob [Y_T S_T^0] = S_0^0$, a risk-neutral measure $\qprob$ is readily defined via the recipe $\ud \qprob = (Y_T S_T^0 / S_0^0)\ud \prob$. However, this is not always the case, as Example \ref{exa: bessel} below will show. Therefore, existence of a stochastic discount factor is a weaker notion than existence of a risk-neutral measure. For some practical applications though, these differences are unimportant. There is further discussion of this point at \S \ref{subsec: SFDs and EMMs} later.

\begin{exa} \label{exa: bessel}
Let $S^0 \equiv 1$ and $S^1$ be a 3-dimensional Bessel process with $S^1_0 = 1$. If $\bF$ is the natural filtration of $S^1$, it can be shown that the \emph{only} stochastic discount factor is $Y = 1 / S^1$, which is a \textsl{strict local martingale} in the terminology of \cite{MR1725406}.
\end{exa}

\subsection{Credit constraints on investment}

In view of the theoretical possibility of continuous trading, in order to avoid so-called \textsl{doubling strategies} (and in order to have the Fundamental Theorem of Asset Pricing holding), \emph{credit constraints} have to be introduced. The wealth of agents has to be bounded from below by some constant, representing the credit limit. Shifting the wealth appropriately, one can assume that the credit limit is set to zero; therefore, only positive wealth processes are allowed in the market.

Since only strictly positive processes are considered, it is more convenient to work with \emph{proportions} of investment, rather than absolute quantities as was the case in Section \ref{sec: SDF_discrete}. Pick some $\bF$-adapted process $\pi = (\pi^1, \ldots, \pi^d)$. For $i =1, \ldots, d$ and $t \in [0, T]$, the number $\pi^i_t$ represents the \emph{percentage} of capital in-hand invested in asset $i$ at time $t$. In that case, $\pi^0 = 1 - \sum_{i=1}^d \pi^i$ will be invested in the savings account. Denote by $X^\pi$ the wealth generated by starting from unit initial capital ($X^\pi_0 = 1$) and invest according to $\pi$. Then,
\begin{equation} \label{eq: wealth dynamics, Ito}
\frac{\ud X^\pi_t}{X^\pi_t} \, = \, \sum_{i=0}^d \pi^i_t \frac{\ud S^i_t}{S^i_t} = \pare{ r_t + \inner{\pi_t }{b_t - r_t \fone}} \ud t + \inner{\sigma_t \pi_t}{\ud W_t}.
\end{equation}
To ensure that the above wealth process is well-defined, we must assume that
\begin{equation} \label{eq: feasible portfolios Ito}
\int_0^T |\inner{\pi_t }{b_t - r_t \fone}| \, \ud t < + \infty \text{ and } \int_0^T \inner{\pi_t}{c_t \pi_t}  \ud t < + \infty, \, \prob \text{-a.s.}
\end{equation}
The set of all $d$-dimensional, $\bF$-adapted processes $\pi$ that satisfy \eqref{eq: feasible portfolios Ito} is denoted by $\Pi$. A simple use of the integration-by-parts formula gives the following result.

\begin{prop}
If $Y$ is a stochastic discount factor, then $Y X^\pi$ is a local martingale for all $\pi \in \Pi$.
\end{prop}

\subsection{Connection with ``no free lunch'' notions}

The next line of business is to obtain an \emph{existential} result about stochastic discount factors in the present setting, also connecting their existence to a no-arbitrage-type notion. Remember from \S \ref{subsec: log-utility} the special stochastic discount factor that is the reciprocal of the log-optimal wealth process. We proceed somewhat heuristically to compute the analogous processes for the It\^o-process model. The linear stochastic differential equation \eqref{eq: wealth dynamics, Ito} has the following solution, expressed in logarithmic terms:
\begin{equation} \label{eq: log-wealth}
\log X^\pi = \int_0^\cdot \pare{ r_t + \inner{\pi_t }{b_t - r_t \fone} - \frac{1}{2}  \inner{\pi_t}{c_t \pi_t}} \ud t + \int_0^\cdot \inner{\sigma_t \pi_t}{\ud W_t}
\end{equation}
Assuming that the local martingale term $ \int_0^\cdot \inner{\sigma_t \pi_t}{\ud W_t}$ in \eqref{eq: log-wealth} is an actual martingale, the aim is to maximize the expectation of the drift term. Notice that we can actually maximize the drift \emph{pathwise} if we choose the portfolio $\pi_* = c^{-1} (b - r \fone)$. We need to ensure that $\pi_*$ is in $\Pi$. If is easy to see that \eqref{eq: feasible portfolios Ito} are both satisfied if and only if $\int_0^T |\lambda^*_t|^2 \ud t < \infty$ $\prob$-a.s., where $\lambda^* := \sigma c^{-1} (b - r \fone)$ is the special risk-premium of Proposition \ref{prop: market prices of risk}. Under this assumption, $\pi_* \in \Pi$. Call $X^* = X^{\pi_*}$ and define
\begin{equation} \label{eq: log disc factor Ito}
Y^* \, := \, \frac{1}{X^*} \, = \, \beta \exp \pare{- \int_0^\cdot \inner{\lambda^*_t}{\ud W_t} - \frac{1}{2} \int_0^\cdot |\lambda_t^*|^2 \ud t}.
\end{equation}
Use the integration-by-parts formula it is rather straightforward to check that $Y^*$ is a stochastic discount factor. In fact, the ability to define $Y^*$ \emph{is} the way to establish that a stochastic discount factor exists, as the next result shows.

\begin{thm}
For the It\^o process model considered above, the following are equivalent.
\begin{enumerate}
  \item The set of stochastic discount factors is non-empty.
  \item $\int_0^T |\lambda^*_t|^2 \ud t$, $\prob$-a.s.; in that case, $Y^*$ defined in \eqref{eq: log disc factor Ito} is a stochastic discount factor.
  \item For any $\epsilon > 0$, there exists $\ell = \ell(\epsilon) \in \Real_+$ such that $\prob[X_T^\pi > \ell] < \epsilon$ uniformly over all portfolios $\pi \in \Pi$.
\end{enumerate}
\end{thm}

The property of the market described in statement (3) of the above Theorem is coined \textsl{No Unbounded Profit with Bounded Risk} in \cite{MR2335830}, where the interest reader is referred to.

\smallskip

The next \emph{structural} result about the stochastic discount factors in the It\^o process setting revels the importance of $Y^*$ as a building block.

\begin{thm} \label{thm: structure of SDFs}
Assume that $\bF$ is the filtration generated by the Brownian motion $W$. Then, any stochastic discount factor $Y$ in the previous It\^o process model can be decomposed as $Y = Y^* \, N^\kappa$, where $Y^*$ was defined in \eqref{eq: log disc factor Ito} and
\[
N^\kappa_t = \exp \pare{ - \int_0^t \inner{\kappa_u}{\ud W_u} - \int_0^t |\kappa_u|^2 \ud u }, \quad t \in [0, T]
\]
where $\kappa$ is a $m$-dimensional $\bF$-adapted process with $\sigma^\top \kappa = 0$.
\end{thm}

If the assumption that $\bF$ is generated by $W$ is removed, one still obtains a similar result with $N^\kappa$ being replaced by \emph{any} positive $\bF$-martingale $N$ with $N_0 = 1$ that is \emph{strongly} orthogonal to $W$. The specific representation obtained in Theorem \ref{thm: structure of SDFs} comes from the \emph{Martingale Representation Theorem} of Brownian filtrations; check for example \cite{MR1121940}.

\subsection{Stochastic discount factors and equivalent martingale measures} \label{subsec: SFDs and EMMs}

Consider an agent that uses a stochastic discount factor $Y$ for valuation purposes. There is a possibility that $Y S^i$ could be a \emph{strict local} $\prob$-martingale for some $i=0, \ldots, d$, which would mean that\footnote{The inequality follows because positive local martingales are supermartingales --- see for example \cite{MR1121940}.} $S_0^i > \expec^{\prob} [Y_T S_T^i]$. The last inequality is puzzling in the sense that the agent's indifference price for the $i$th asset, which is $\expec^{\prob} [Y_T S_T^i]$, is \emph{strictly} lower than the market price $S_0^i$. In such a case, the agent would be expected to wish to short some units of the $i$th asset. This is indeed what is happening; however, because of credit constraints, this strategy is infeasible. Example \ref{exa: bessel 2} below will convince you of this fact. Before presenting the example, an important issue should be clarified. One would rush to state that such ``inconsistencies'' are tied to the notion of a stochastic discount factor as it appears in Definition \ref{defn: stoch disc fact Ito}, and that is is \emph{strictly} weaker than existence of a probability $\qprob \sim \prob$ that makes all discounted processes $\beta S^i$ \emph{local} $\qprob$-martingales for $i = 0, \ldots, d$. Even if such a probability \emph{did} exist, $\beta S^i$ could be a \emph{strict local} $\qprob$-martingale for some $i=1, \ldots, d$; in that case, $S_0^i > \expec^{\qprob} [\beta_T S_T^i]$ and the same mispricing problem pertains.

\begin{exa} \label{exa: bessel 2}
Let $S^0 \equiv 1$, $S^1$ be the \emph{reciprocal} of a 3-dimensional Bessel process starting at $S^1_0 = 1$ under $\prob$ and $\bF$ be the filtration generated by $S^1$. Here, $\prob$ is the \emph{only} equivalent local martingale measure and $1 = S^1_0 > \expec^\prob [S_T^1]$ for all $T > 0$. This is a \emph{complete} market --- an agent can start with capital $\expec^\prob [S_T^1]$ and invest in a way so that at time $T$ the wealth generated is exactly $S_T$. Naturally, the agent would like to long as much as possible from this replicating portfolio and short as much as possible from the actual asset. However, in doing so, the possible downside risk is infinite throughout the life of the investment and the enforced credit constraints will disallow for such strategies.
\end{exa}

In the context of Example \ref{exa: bessel 2}, the law of one price fails, since the asset that provides payoff $S^1_T$ at time $T$ has a market price $S^1_0$ and a replication price $\expec^\prob[S^1_T] < S^1_0$. Therefore, if the law of one price is to be valid in the market, one has to insist on existence of an equivalent (\emph{true}) martingale measure $\qprob$, where each discounted process $\beta S^i$ is a \emph{true} (and not only local) $\qprob$-martingale for all $i=0, \ldots, d$. For pricing purposes then, it makes sense to ask that the stochastic discount factor $Y^\kappa$ that is chosen according to Theorem \ref{thm: structure of SDFs} is such that $Y^\kappa S^i$ is a true $\prob$-martingale for all $i=0, \ldots, d$. Such stochastic discount factors give rise to probabilities $\qprob^\kappa$ that make all deflated asset-price-process $\qprob^\kappa$-martingales and can be used as pricing measures.

Let us specialize now to a Markovian model where the interest rate process $(r_t)_{t \in [0, T]}$ is deterministic and the process $\sigma$ is a funnction of the assets, i.e., $\sigma = \Sigma(S)$, where $\Sigma$ is a deterministic function from $\Real^d$ to the space of $(m \times d)$-matrices. As long as a claim written only on the traded assets is concerned, the choice of $\qprob^\kappa$ for pricing is irrelevant, since the asset prices under $\qprob^\kappa$ have dynamics
\[
\frac{\ud S^i_t}{S^i_t} = r_t \ud t + \inner{\Sigma^i(S_t)}{\ud W^{\kappa}_t}, \quad t \in [0, T], \quad i=1, \ldots, d,
\]
where $W^\kappa$ is a $\qprob^\kappa$-Brownian motion; in particular, the process $(S)_{t \in [0, T]}$ has the same law under any $\qprob^\kappa$.
However, if one is interested on pricing a claim written on a non-traded asset whose price process $Z$ has $\prob$-dynamics
\[
\ud Z_t = a_t \ud t + \inner{f_t}{\ud W_t}, \quad t \in [0, T]
\]
for $\bF$-adapted $a$ and $f=(f^1, \ldots, f^m)$, then the $\qprob^\kappa$-dynamics of $Z$ are
\[
\ud Z_t = \left( a_t - \inner{f_t}{\lambda^*_t} - \inner{f_t}{\kappa_t} \right) \ud t + \inner{f_t}{\ud W^\kappa_t}, \quad t \in [0, T]
\]
The dynamics of $Z$ will be independent of the choice of $\kappa$ only if the volatility structure of the process $Z$, given by $f$, is in the range of $\sigma^\top = \Sigma(S)^\top$. This will mean that $\inner{f}{\kappa} = 0$ for all $\kappa$ such that $\sigma^\top \kappa = 0$ and that $Z$ is perfectly replicable using the traded assets. As long st there is any randomness in the movement in $Z$ that cannot be captured by investing in the traded assets, i.e., if there exists some $\kappa$ with $\sigma^\top \kappa = 0$ and $\inner{f}{\kappa}$ not being identically zero, perfect replicability fails and pricing becomes a more complicated issue, depending on the preferences of the particular agent as given by the choice of $\kappa$ to form the stochastic discount factor.

\bibliographystyle{siam}
\bibliography{stoch_disc_fact}
\end{document}